\documentclass[letterpaper, 10 pt, conference]{ieeeconf}  

\IEEEoverridecommandlockouts                              

\usepackage{amsmath} 
\usepackage{booktabs}
\usepackage{graphicx}
\usepackage{makecell}
\usepackage{xcolor}

\usepackage{etoolbox}
\makeatletter
\patchcmd{\@makecaption}
  {\scshape}
  {}
  {}
  {}
\makeatother

\usepackage{url}

\title{\LARGE \bf
The ``I Don't Know" Filter: \\Enhancing Agentic Reliability in Function Calling
}

\author{Stefan Broecker$^{1}$, Mason del Rosario$^{2}$, Boris Selitser$^{2}$, Thomas Strohmer$^{3}$  
\thanks{$^{1}$University of California, Davis, Department of Computer Science,
        {\tt\small sbroecker@ucdavis.edu}}%
\thanks{$^{2}$Okareo, Inc.,
        {\tt\small \{mason, boris\}@okareo.com}}%
\thanks{$^{3}$University of California, Davis, Department of Mathematics,
        {\tt\small strohmer@math.ucdavis.edu}}%
}
\begin{document}

\maketitle
\thispagestyle{empty}
\pagestyle{empty}

\begin{abstract}

The language models that underpin agents have seen a rapid rise in performance on function calling benchmarks. However, the metrics used in the training and evaluation of these models often encourage models to make positive claims even when the answer is uncertain, leading to hallucinations. Such hallucinations can be disastrous when language models are trusted to use function calls to make decisions in high stakes applications. To that end, we propose an agent evaluation metric that takes into account the negative outcomes associated with incorrect function calls. Further, to catch hallucinations before they can cause harm, we propose a lightweight trainable filter that can quantify a language model's uncertainty and remove potentially harmful function calls. By training that filter to detect and suppress uncertain function calls without modifying the underlying model, we demonstrate a practical path toward agents that know when to say ``I don't know," a property we argue is essential to production reliability.

\end{abstract}

\section{INTRODUCTION}

Agents are language models (LMs) that combine tools, memory, and reasoning to perform tasks \cite{Wang2024}. As agents have improved at tool use, developers in critical industries such as digital infrastructure \cite{hatami2026securing}, health care, and public services \cite{Deloitte2026StateOfAI} have begun to trust agents to make decisions in production settings. As an example, errant tool calls in such production environments have resulted in airfares offered at disallowed discounts and eggs purchased at exorbitant prices \cite{riedl2025ai}. 

A large scale study of agents in production showed a persistent gap in agent performance on benchmarks and in production systems \cite{pan2026measuringagentsproduction}. In that study, reliability was most commonly cited as the bottleneck to agent deployment. A key part of that reliability gap is hallucinations, and given agents’ ability to take autonomous action in real-world decision making, hallucinations can cause much more severe and material damage than standalone LMs.

\cite{kalai2025languagemodelshallucinate} makes the claim that hallucinations persist because of conventions in LM evaluations. In most evaluations, an incorrect answer is weighed the same as a non-answer, so guessing when uncertain improves accuracy. However, Bastounis et al. state that this tendency to provide a non-answer when uncertain is a fundamental roadblock to ``trustworthy" (i.e., reliable) AI, and they describe this tendency to hallucinate and the inability to state ``I don't know" as a critical component of the Consistent Reasoning Paradox \cite{bastounis2024consistentreasoningparadoxintelligence}.

In this work, we study ``agent reliability" through this lens: the agent's ability to express uncertainty, i.e., by explicitly stating ``I don't know" (IDK). Towards this end, we make the following contributions:

\begin{itemize}
    \item An accuracy metric that penalizes incorrect function calls, emphasizing the danger of an autonomous agent guessing at an answer instead of admitting uncertainty and encouraging IDK responses
    \item A simple filter that uses ``whitebox," ``graybox," and ``blackbox" features to identify likely-incorrect function calls (cases of IDK) to reduce the likelihood of hallucinations
    \item A simple method for generating synthetic data that can be used to train that filter, enabling deployment even when labeled function calling data is limited
    \item A study of the filter's effectiveness in reducing uncertainty across different open source LMs and function calling benchmarks
\end{itemize}

With these contributions, we demonstrate that explicitly measuring an agent's uncertainty and modulating its output accordingly can reduce the likelihood of incorrect function calls, improving its reliability.

\section{BACKGROUND}

\subsection{Uncertainty quantification for LMs}

Existing work has applied uncertainty quantification to measure the specific sources of uncertainty in the output of LMs (\cite{10.1145/3711896.3736569, 10.1145/3744238, lin2024generatingconfidenceuncertaintyquantification}). These works have mostly geared towards hallucination detection in freeform, linguistic LM responses rather than function calls specifically. Works in this field use various metrics to quantify the variability of LM outputs. Of note for the present work, Li et al. calculate a so-called ``semantic volume," or the volume of the parallelepiped formed by the embeddings of outputs \cite{li2025semanticvolumequantifyingdetecting}. Huang et al. \cite{huang2023look} use a formulation of variation ratio, adapted by Wang et al. \cite{wang2022exploratorystudyairisk} to be used on LM outputs, to measure the statistical dispersion of outputs.

In work more specifically focused on function calling, Rabanser et al. \cite{rabanser2026scienceaiagentreliability} break down agent reliability into four dimensions: consistency (repeatable behavior across runs), robustness (stability under input and environmental perturbations), predictability (calibrated confidence and discrimination of correct/incorrect predictions), and safety (bounded severity when failures occur).

While we claim that this work has beneficial implications in all of these domains, our methods focus on \textbf{consistency} across different sampling runs. More specifically, we investigate an agent's consistency under ``prediction uncertainty'', i.e. the uncertainty in an LM's output given an identical input that arises from both variability in the training data (i.e., aleatoric uncertainty) and in genuine gaps in the LM's knowledge (i.e., epistemic uncertainty) \cite{10.1145/3711896.3736569}.

\subsection{Reliable function calling for agents}

Existing work on function calling focuses on improving the standard accuracy measures that do not account for an LM's uncertainty. Further, when dealing with agent reliability, work in this area focuses on \textbf{robustness}, or the ability to generate the correct function call under input perturbations \cite{rabinovich-anaby-tavor-2025-robustness, wang2023assessingreliabilitylargelanguage}.

A work with more direct relevance to the current work is $\tau$-bench \cite{yao2024taubenchbenchmarktoolagentuserinteraction}, a benchmark which aims to measure \textbf{consistency}. $\tau$-bench uses a metric of agent reliability over repeated trials on the same task which Yao et al. call pass\string^k. They define pass\string^k as the chance that all k
i.i.d. task trials are successful, averaged across tasks. While pass\string^k is a valid indicator of LM reliability, the metric does not penalize highly confident yet incorrect answers (i.e., hallucinations), a shortcoming which we address with our proposed metric.

\subsection{Language model features}

To train a filter that measures an LM's uncertainty on a given input text, we need to select a feature set. In this study, we consider features of varying ``invasiveness'', which we call \textbf{whitebox}, \textbf{graybox}, or \textbf{blackbox}.

In this work, we opt for small language models (SLMs) that allow access to \textbf{whitebox} features, or the internal state of the LM (i.e., model parameters, activation values, etc.). In \cite{marks2024geometrytruthemergentlinear}, Marks et al. show that internal states can reliably encode things like truthfulness, and in \cite{orgad2025llmsknowshowintrinsic}, Orgad et al. use the internal states of LMs to detect hallucinations.

While it is natural to question the performance of SLMs relative to large language models (LLMs), there is evidence to suggest that SLMs can be equally or more powerful than LLMs when the ``objective is schema- and API-constrained accuracy rather than open-ended generation" \cite{sharma2025smalllanguagemodelsagentic}.

While having access to whitebox features is ideal, many production settings only allow the use of \textbf{blackbox} features of LMs (i.e., the generated text outputs). Occasionally, such API-gated LMs also provide access to their log probabilities (``logprobs''), and since these logprobs offer a glimpse of the LM's internal state at its outer boundary, we refer the features derived from them as \textbf{graybox} features.

In this study, we consider using the full range of whitebox, graybox, and blackbox features in our feature selection for the uncertainty filter, and we analyze the relative importance of each in measuring the LM's uncertainty.


\subsection{Our Work}

Our work builds on two of the key observations mentioned in this section. First, uncertainty quantification research demonstrates that variability in LM outputs can serve as a reliable signal of model uncertainty. Second, existing function calling benchmarks, while measuring accuracy and consistency, do not penalize incorrectness, leaving a gap between benchmark performance and the reliability demands of production agents. We bridge these ideas by treating output inconsistency across repeated function calls as a proxy for agent uncertainty, and by training a classifier to act on that signal by filtering out likely-incorrect calls and replacing them with explicit abstentions. The following section describes this approach in detail.

\section{METHODS}

Identifying when an LM agent should abstain from answering a question rather than guessing is central to improving the reliability of agentic systems. To that end, we introduce two concepts: a new evaluation metric and a trained classifier filter. We then validate both across multiple benchmarks and LM agents. The sections below describe the metric, the classifier design, and the experiments used to evaluate them\footnote{All code and prompts used in this manuscript can be found at https://github.com/SBroecker/idkscore.}. 

\subsection{IDK Score}

To better quantify the effectiveness of our classifier in improving the reliability of an agent, we introduce the \textit{I Don't Know Score} (IDKS). IDKS takes into account the negative effect caused by producing an incorrect function call. We define it as: \[IDKS = \frac{C - I}{N},\] where $C$ is the number of correct function calls, $I$ is the number of incorrect function calls, and $N$ is the total number of function calls. IDKS ranges from -1 to 1, where -1 means every function call is incorrect and 1 means that every function call is correct. Notably, IDKS does not include agent outputs that indicate uncertainty, like ``I don't know." These abstentions are counted as neutral in the calculation of IDKS. Further, we acknowledge that the choice of -1 as the penalty for an incorrect function call in IDKS is a simplification that practitioners can adjust based on their use case.

With this metric in hand, we can now describe the filter designed to improve it.

\subsection{Classifier Design}

\begin{figure}[ht]
  \centering
  \includegraphics[width=\linewidth]{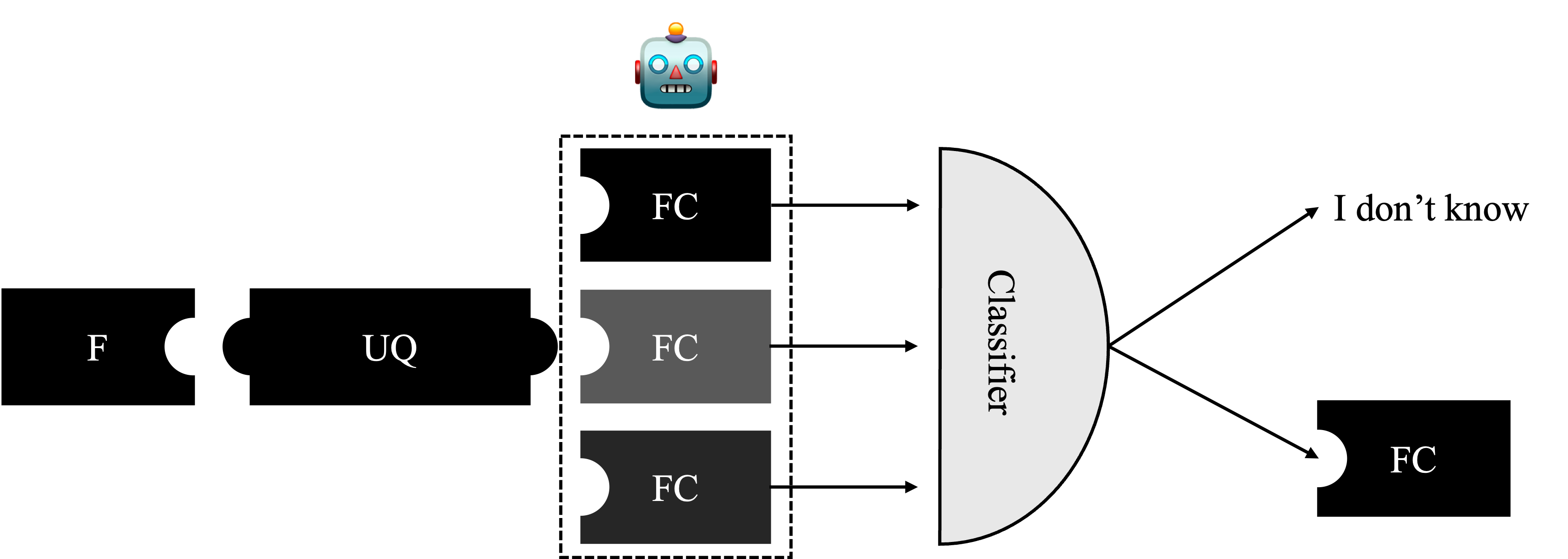}
  \caption{A classifier being used as a filter for an LM agent. Given the same input functions (F) and user question (UQ), an agent generates multiple function calls (FC). Those repeated function calls are fed into the classifier, which decides whether to use the output or filter it out and respond with "I don't know".}
  \label{fig:classifier_filter}
\end{figure}

The core of our approach is a classifier trained to distinguish reliable agent outputs from uncertain ones using features derived from repeated generations. At inference time, if the classifier's uncertainty estimate for a given set of outputs exceeds a threshold, the agent's output is filtered out and replaced with an abstention. This concept is illustrated in Figure \ref{fig:classifier_filter}.

\subsection{Experimental Setup}

\subsubsection{Models}

We evaluate our classifier filter using three open-source SLMs that can be run locally using Huggingface's Transformers library \cite{wolf2020huggingfacestransformersstateoftheartnatural}, enabling access to whitebox and graybox features: Microsoft's \textit{Phi-4-mini-instruct}, Alibaba Cloud's \textit{Qwen2.5-3B-Instruct}, and Meta's \textit{Llama-3.2-3B-Instruct} \cite{microsoft2025phi4minitechnicalreportcompact, qwen2.5, grattafiori2024llama3herdmodels}. These models were selected because they contain a similar number of parameters but span a range of model families, allowing us to assess the filter's  portability across architecturally distinct agents. Llama and Qwen models are used with default decoding parameters. For Llama this means a temperature of 0.9 and top\_p value of 0.6. For Qwen this means a temperature of 0.7, a top\_p value of 0.8, and a top\_k value of 20. Since Phi models use greedy decoding by default, and since a greedy decoding strategy doesn't produce any output variability when given the same inputs, we use Llama's default decoding parameters for the Phi model.

\subsubsection{Training Data}

We train our classifier on a dataset consisting of synthetic user questions. Our data are created using several Berkeley Function Calling Leaderboard (BFCL) datasets as a starting point \cite{patil2023gorilla}. Specifically, we use BFCL's Simple Python (single functions and function calls), Parallel (single functions with multiple expected function calls), Multiple (multiple possible functions with a single expected function call), and Parallel Multiple (multiple possible functions with multiple expected function calls). While we use BFCL as our starting point, we note that our methodology is generalizable and could thus be used with any existing function calling dataset.

Each datapoint in the BFCL datasets consists of: available functions and their definitions, a user question, and expected function calls. We use that same structure for all of our data. To generate synthetic data, for each datapoint in the aforementioned BFCL datasets, we prompt OpenAI's \textit{gpt-4.1-mini} to create a new user question that would lead to the possible answers provided in the datapoint given the available functions. This is done five times for each datapoint to create five unique variations of user questions.

\begin{figure}[ht]
  \centering
  \includegraphics[width=\linewidth]{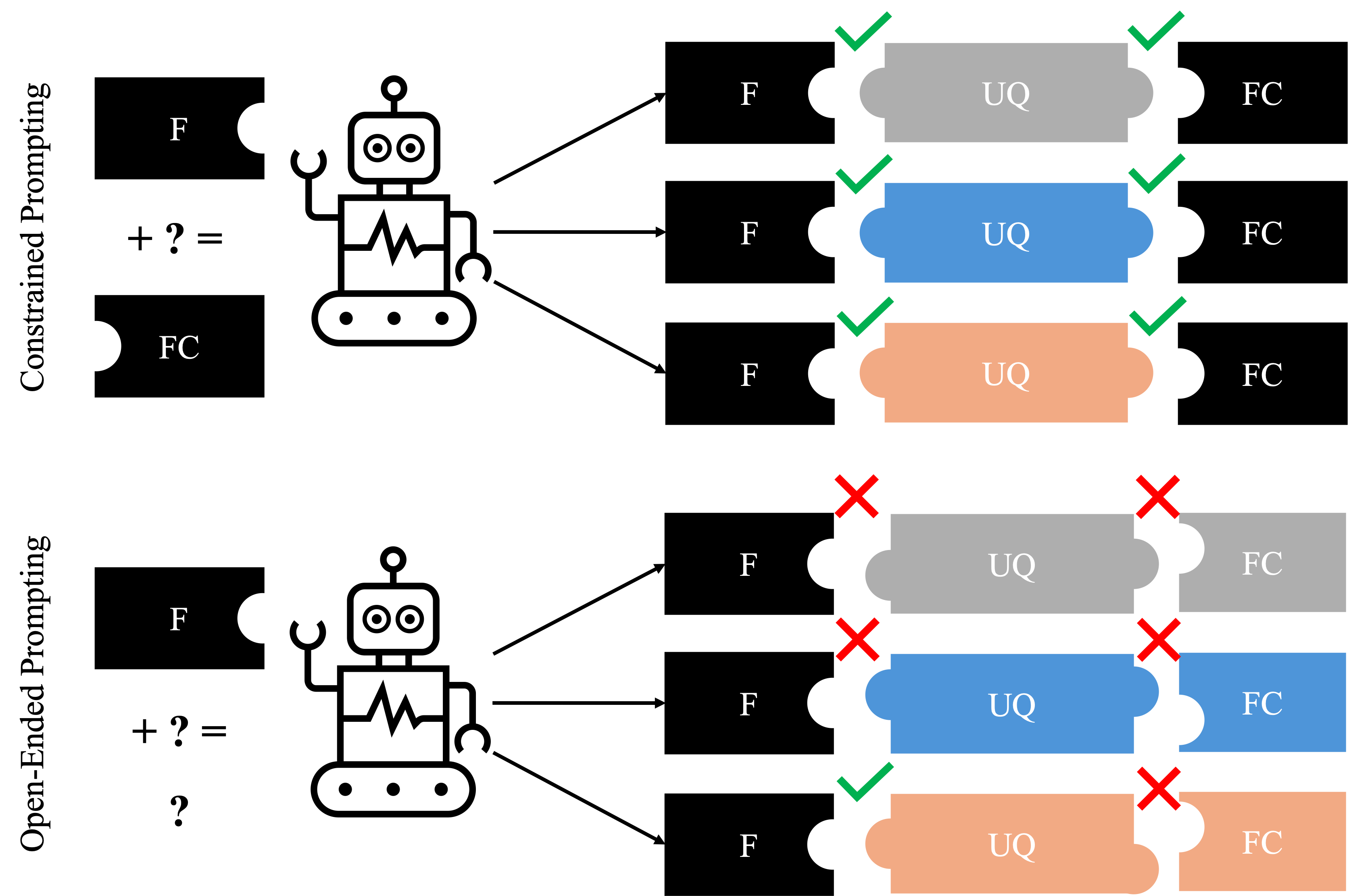}
  \caption{Synthetic data is generated using constrained prompting, where an LLM is given functions (F) and the expected function calls (FC) and is expected to create valid user questions (UQ). In contrast, open-ended prompting, where the LLM is given less predefined information, produces less reliable function calling scenarios, where the functions, questions, and function call sometimes don't agree with each other.}
  \label{fig:data_generation}
\end{figure}

We call this prompting approach constrained prompting, which we compare to more open-ended prompting, where, for example, an LLM might be prompted to generate both a question and possible function calls given a set of available functions. We found that open-ended prompting produced outputs where the generated questions frequently failed to align with the expected function calls, making them unsuitable as a training data source. An illustrated overview of our constrained prompting approach and an alternative open-ended prompting approach can be seen in Figure \ref{fig:data_generation}.

Having established the training set, we now describe the held-out data used to evaluate classifier performance.

\subsubsection{Validation Data}

We use the aforementioned single-shot BFCL datasets as validation in our experiments: Simple Python, Parallel, Multiple, and Parallel Multiple. We additionally use Simple Java (single functions and function calls with Java formatting), Simple JavaScript (single functions and function calls with JavaScript formatting), and Irrelevance (none of the function choices provided are relevant to the user query and none should be invoked). We also use a subset of datapoints from APIGen, a dataset of synthetically generated function calling examples \cite{liu2024apigenautomatedpipelinegenerating}. To create that subset, we clean it and filter it, and then reformat it to match the format of BFCL datapoints. We then take a random subset of 10,000 examples.


\subsubsection{Feature Extraction} \label{sec:features}

Features for the classifier are created using the differences observed over repeated generations by the agent. Three different levels of features are extracted. In order of decreasing invasiveness for the LM agent, features are extracted from the final internal state of the agent (whitebox features), the logprobs during generation (graybox features), and the text that was generated (blackbox features). For blackbox features, downstream features are calculated after the text is embedded using the text embedding model trained by \cite{liu2024codexembed}.

Features are generated from raw whitebox and blackbox features using similar techniques. Namely, semantic volume and variation ratio, two measures of the spread of the features across repeats, are calculated from raw feature vectors \cite{li2025semanticvolumequantifyingdetecting, huang2023look}. Statistical features are then extracted from the pairwise normalized distances between agent responses. Further statistical features, such as minimum probability and entropy, are generated from raw graybox features.

Labels for every datapoint are created using the function call validation tools provided by \cite{patil2023gorilla}. Since groups of function calls are used for training, we use a variation of the pass\string^k metric defined by \cite{yao2024taubenchbenchmarktoolagentuserinteraction} to label our data, where a group of responses is considered valid if and only if every response in the group is valid.

Before classifier training, irrelevant features are filtered out based on a calculation of the mutual information between each feature and the label. Features are further filtered during training using recursive feature elimination. Both of these filtering steps are performed with tools provided by \cite{scikit-learn}. The trained classifier is a random forest model.

\begin{figure}[ht]
  \centering
  \includegraphics[width=0.8\linewidth]{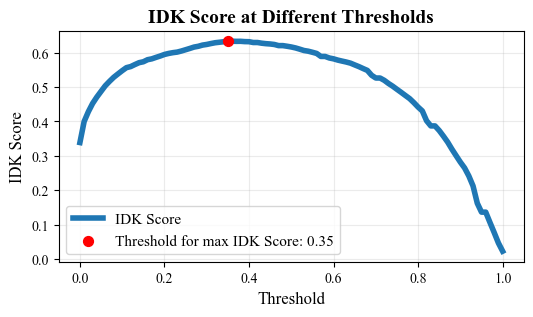}
  \caption{Average performance of a Llama agent with a classifier filter applied at different decision thresholds. Performance suffers at low thresholds because not enough function calls are filtered and at high thresholds because too many function calls are filtered.}
  \label{fig:thresholds}
\end{figure}

The decision threshold for the classifier, above which an agent output is filtered and replaced with an abstention, is selected by maximizing IDKS across benchmarks on the training set. Figure \ref{fig:thresholds} shows the relationship between IDKS and the decision threshold, demonstrating the tradeoff between too permissive and too restrictive of a filter. While optimizing on training data risks overfitting the threshold, we deliberately avoid using validation data for this selection to prevent leakage into our held-out results.

\subsection{Experiments}

We run three main sets of experiments to verify the effectiveness of our filter design.

\subsubsection{Effect of the Number of Repeated Calls}

First, we run experiments varying the number of repeated function calls that are used to build features, ranging from two repeats to five repeats. 

\subsubsection{Effect of Feature Invasiveness}

Next, we run experiments varying the level of invasiveness of features. In increasing level of invasiveness, we use: blackbox features only, blackbox and graybox features, and then blackbox, graybox, and whitebox features.

\subsubsection{Cross-LM Generalizability}

Finally, we run experiments to test the generalizability of features by training a classifier on the outputs of one LM agent and applying that classifier to the outputs of another LM agent.

\section{RESULTS}

We evaluate our classifier filter across three axes: how performance scales with the number of repeated calls used to build features, how much each additional level of feature invasiveness contributes to model performance, and whether a classifier trained on one LM's outputs transfers to another. In each case, performance is reported using IDKS, where a higher score reflects a more reliable agent.

\subsection{Effect of the Number of Repeated Calls}

\begin{figure}[ht]
  \centering
  \includegraphics[width=\linewidth]{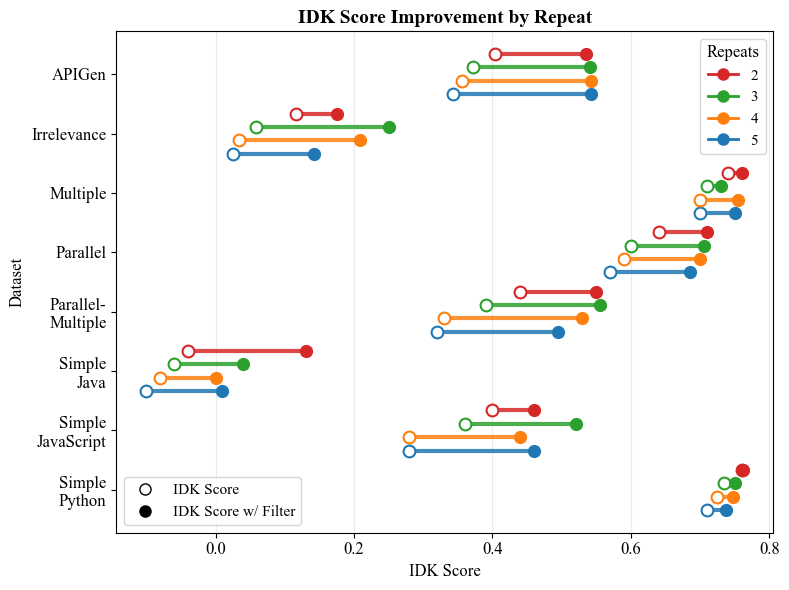}
  \caption{Llama agent performance across all validation datasets, as measured by IDK Score, with and without a trained classifier acting as a filter. Results are grouped along the y-axis according to validation dataset. Within a group, results are colored by the number of repeated function calls. Each colored line represents the difference in IDK Score between the agent's performance without a classifier filter (white dots) and with a classifier filter (colored dots). Longer lines indicate a larger performance gap.}
  \label{fig:repeats}
\end{figure}

Experiments varying the number of repeats show two trends. First, as observed by \cite{yao2024taubenchbenchmarktoolagentuserinteraction}, agents are more likely to produce an incorrect function call as you increase the number of function call repeats. This is indicated by a general decrease in baseline IDKS as the number of repeats increases, as can be seen by the white circles in Figure \ref{fig:repeats}. Second, the boost in performance created by the trained classifier acting as a filter often increases as you increase the number of repeats, as shown by the solid circles in the figure. However, in a number of the benchmarks, including APIGen and BFCL's Parallel dataset, the boost in performance caused by increasing the number of function call repeats is offset by the degradation of the model's baseline performance, creating relatively constant performance across repeats. 

\subsection{Effect of Feature Invasiveness}

Experiments varying the invasiveness of features indicate a boost of performance for each step up in model invasiveness. Table \ref{tab:progressive_features} shows the results of these experiments. Across most validation datasets, performance increases as you move from blackbox to blackbox and graybox and then to blackbox, graybox, and whitebox features. BFCL's Parallel, Parallel Multiple, and Simple JavaScript, are exceptions to this, where the addition of whitebox features hurts performance slightly. BFCL's Simple Java is another exception, where performance decreases as gray- and whitebox features are added. An apparent outlier, we believe this result is a consequence of the Llama agent's general poor performance on Simple Java and a subsequent lack of a reliable signal.

\begin{table}
\centering
\begin{tabular}{lcccc}
\toprule
 & Baseline & Black & \makecell{Black +\\ Gray} & \makecell{Black +\\Gray + \\ White} \\
\midrule
APIGen & 0.342 & 0.453 & 0.532 & 0.541 \\
Irrelevance & 0.025 & 0.042 & 0.217 & 0.229 \\
Multiple & 0.700 & 0.725 & 0.750 & 0.760 \\
Parallel & 0.570 & 0.625 & 0.680 & 0.670 \\
Parallel Multiple & 0.320 & 0.465 & 0.505 & 0.490 \\
Simple Java & -0.100 & 0.050 & 0.000 & -0.010 \\
Simple JavaScript & 0.280 & 0.360 & 0.400 & 0.380 \\
Simple Python & 0.710 & 0.685 & 0.738 & 0.743 \\
\bottomrule
\end{tabular}
\caption{IDK Score of a Llama agent trained on Llama outputs with varying levels of feature invasiveness. Performance is shown on different validation datasets. Black, gray, and white refer to blackbox, graybox, and whitebox features, respectively, as described in Section \ref{sec:features}.}
\label{tab:progressive_features}
\vspace{-20pt}
\end{table}

\subsection{Cross-LM Generalizability}

\begin{figure}[ht]
  \centering
  \includegraphics[width=\linewidth]{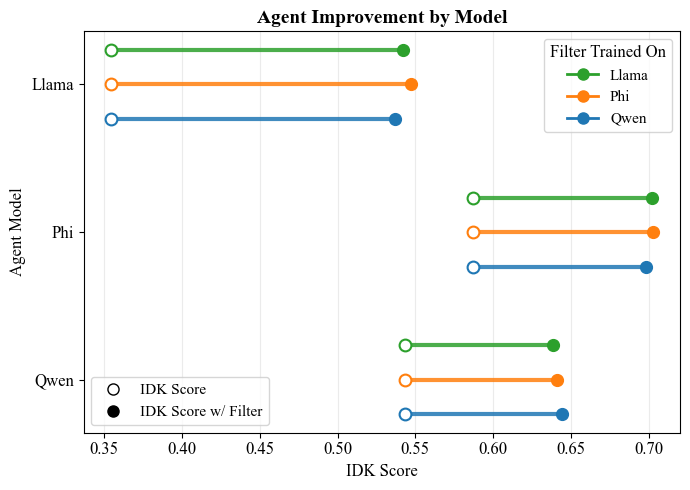}
  \caption{IDK Score for Llama, Phi, and Qwen agents averaged across all validation datasets, with and without a trained classifier acting as a filter. Results are grouped along the y-axis according to the agent model. Within a group, colors represent the model that was used to train the classifier filter. White dots represent the agent's performance without the filter and colored dots represent the agent's performance with a filter. Consistent performance gaps indicate that features are able to generalize across agent models.}
  \label{fig:overview_result}
\end{figure}

Experiments across LM agents indicate the portability of the features used to train our classifiers. Figure \ref{fig:overview_result} shows the averaged performance across all validation datasets of each LM agent with and without a classifier filter. Further, it shows the agent's performance when the filter is trained on features created by its own LM and different LMs. Performance is strongest for Phi and Qwen agents when the classifier is trained on the LM's own features, with performance slightly dropping when the classifier is trained on a different LM's features. The Llama agent is the exception, where a filter trained on Phi features is the strongest. Further work could explore whether this is a fluke or whether Phi's outputs are a particularly informative signal.

\section{DISCUSSION}

Our results demonstrate that a consistency-based classifier filter is an effective mechanism for improving the reliability of LM agents in function calling settings. Across benchmarks and model families, applying the filter yields consistent gains in IDKS, confirming that output inconsistency across repeated generations is a useful proxy for agent uncertainty. Crucially, the portability of classifiers across LM agents, where performance drops only slightly when a classifier trained on one model's outputs is applied to another, suggests that the uncertainty signal captured by our features reflects general properties of function calling difficulty rather than idiosyncrasies of any single model.

\subsection{Consistency Trends Relative to Prior Work}

Our results also shed light on how agent consistency has improved since \cite{yao2024taubenchbenchmarktoolagentuserinteraction}. In that work, repeated sampling led to rapid degradation in pass\string^k, with model performance dropping by almost half by the fourth repeat for some models. In our experiments, we observe a meaningfully slower degradation rate of $\sim$10\% by the fifth repeat for most benchmarks, suggesting that the LMs used in this study have become more consistent in their function calling behavior. However, our results show that even this reduced degradation rate is not negligible: baseline IDKS still declines as the number of repeated calls increases (Figure \ref{fig:repeats}), reinforcing that consistency-based filtering remains a valuable intervention even for more capable models.

\subsection{Failure Modes}

Despite its effectiveness, the filter has two notable failure modes that stem from the fundamental limitation of using inconsistency as a proxy for uncertainty.

The first is the case of confident incorrectness: when a model consistently produces the same wrong function call across all repeated generations. In this case, the classifier observes low variance and incorrectly treats the output as reliable. One observed example of this failure case is the agent being asked to find the average of a list of floats and producing a function call for a list of ints across all five repeated function calls. In such cases, the model's confidence is high but misplaced, and our filter has no signal with which to intervene. Addressing this failure mode likely requires complementary approaches, such as external type verification.

The second failure mode is superficial inconsistency due to argument formatting variation: when a model produces semantically identical function calls that differ only in surface-level formatting. For example, $3x\string^2$, $3*x\string^2$, and $3*x**2$ may be treated as meaningful uncertainty by the classifier  and thus be incorrectly filtered out even if they all mean the same thing. One remedy would be to apply lightweight normalization to generated function calls before feature extraction (e.g., canonicalizing argument types), which we leave as a direction for future work.

\subsection{Implications of Feature Invasiveness}

The progressive improvement in IDKS as features move from blackbox to whitebox (Table \ref{tab:progressive_features}) reflects a natural tradeoff between deployment practicality and classification performance. For practitioners working with API-gated models, our results show that blackbox features alone still yield meaningful gains over the baseline, and that logprobs, where available, provide a meaningful additional signal. Full whitebox access provides the strongest performance, but the marginal gains over blackbox and graybox features are modest on most benchmarks, suggesting that the core uncertainty signal is largely captured at the output and probability levels.

\section{CONCLUSIONS}


We have presented a framework with three main contributions: IDKS, a trained classifier filter, and validation across benchmarks and model families. These results matter because the deployment gap between benchmark performance and production reliability remains one of the primary barriers to adopting agents in high-stakes settings \cite{pan2026measuringagentsproduction}. By explicitly modeling and acting on uncertainty, encouraging agents to say ``I don't know'' rather than guess, our approach targets this gap directly. The framework is also intentionally lightweight, making it amenable to use in production settings. Notably, the classifier is a random forest trained on a modest feature set, and no modification to the underlying LM is required. Further, our use of a generalizable synthetic data generation pipeline demonstrates the feasibility of our approach even in settings where labeled data is sparse.

Several directions remain open for future work. First, principled threshold selection for the classifier is an important practical question that we have not extensively addressed here. The tradeoff between precision and recall in the IDK decision will likely need to be calibrated per-domain. Second, the failure mode of confident incorrectness points to the limits of consistency as a primary proxy for uncertainty. Hybrid approaches that target other forms of reliability (robustness, safety, predictability) may help address this. Finally, while this work focuses on SLMs, extending the framework to closed-source LLMs is a natural next step as agents increasingly rely on frontier models in production.

\addtolength{\textheight}{-12cm}   




\section*{ACKNOWLEDGMENT}

Stefan Broecker and Thomas Strohmer would like to acknowledge support from the following grants: NSF DMS-2208356, NIH R01HL16351, P41EB032840, and DE-SC0023490. The authors would like to acknowledge Zachary del Rosario for providing invaluable feedback on an early version of this manuscript.

\end{document}